\documentclass{LMCS}
\usepackage{amsmath, amssymb, hyperref}

\newcommand\dsp{\enspace} 
\newcommand\eg{\textit{e.g.}}

\renewcommand\emptyset{\varnothing}

\newcommand\KA{\ensuremath{\mathsf{KA}}}
\newcommand\KAT{\ensuremath{\mathsf{KAT}}}

\newcommand\RKA{\ensuremath{\mathsf{RKA}}}

\newcommand\RKAT{\ensuremath{\mathsf{RKAT}}}

\newcommand\imp{\mathbin{\rightarrow}}

\newcommand\Rex{\ensuremath{\mathsf{RExp}}}
\newcommand\RexS{\ensuremath{\Rex_\Sigma}}

\newcommand\Tr{\ensuremath{\mathsf{Tr}}}

\renewcommand\P{\ensuremath{\mathsf{P}}}
\newcommand\B{\ensuremath{\mathsf{B}}}
\newcommand\C{\ensuremath{\mathsf{C}}}
\newcommand\RexPB{\ensuremath{\Rex_{\P,\B}}}

\newcommand\REG{\ensuremath{\mathsf{REG}}}
\newcommand\id{\iota}
\newcommand\Horn[1]{\ensuremath{\mathcal{H}#1}}
\newcommand\complementop{\ensuremath{\overline{\vphantom{t}~~}}}

\newcommand\nats{\mathbb{N}}

\newcommand\rel[1]{2^{#1 \times #1}}

\newcommand\PSPACE{{\it PSPACE}}

\newcommand\con{\ensuremath{\mathrm{CON}}}
\newcommand\nfa{\ensuremath{\mathrm{NFA}}}

\newcommand\ins{\ensuremath{\mathsf{insert}}}

\newcommand{\sigs}{\ensuremath{{\Sigma^*}}}

\newcommand\romanize{\def\theenumi{\roman{enumi}}\def\labelenumi{{\rm(\theenumi)}}}

\newcommand\union{\mathbin{\cup}}

\newcommand\note[1]{} 
\newcommand\mynote[1]{} 
\newcommand\longnote[1]{} 

\def\doi{1 (3:4) 2005}
\lmcsheading%
{\doi}
{15}
{}
{}
{Apr.~20, 2005}
{Dec.~21, 2005}
{}

\begin{document}

\title{Modularizing the Elimination of $r=0$ in Kleene Algebra}
\author[C.~Hardin]{Christopher Hardin}
\address{Department of Mathematics\\
Smith College\\
Northampton, Massachusetts 01063, USA}
\email{chardin@math.smith.edu}


\begin{abstract}
\noindent 
  Given a universal Horn formula of Kleene algebra with hypotheses of
  the form $r=0$, it is already known that we can efficiently
  construct an equation which is valid if and only if the original
  Horn formula is valid.  This is an example of \emph{elimination of
  hypotheses}, which is useful because the equational theory of Kleene
  algebra is decidable while the universal Horn theory is not.  We
  show that hypotheses of the form $r=0$ can still be eliminated in
  the presence of other hypotheses.  This lets us extend any technique
  for eliminating hypotheses to include hypotheses of the form $r=0$.
\end{abstract}

\maketitle

\section{Introduction}

Kleene algebra (\KA) arises in many areas of computer science,
such as automata
theory, the design and analysis of algorithms, dynamic logic, and program
semantics.  Many of these applications are enhanced by using Kleene algebra
with tests (\KAT), which combines \KA\ with Boolean algebra.

We can use \KAT\ to reason propositionally about programs (see
\cite{BK02, KP00} for examples).
The equivalence of an optimized and unoptimized program, the
equivalence of an annotated and unannotated program, and partial correctness
assertions can all be expressed as equations.
The equational theory of \KAT\ is well understood
and has many useful properties; in particular, it is decidable (in
\PSPACE) and the theory remains unchanged when we restrict to relational
interpretations \cite{CKS96,KS96}.
(Relational interpretations are of the greatest
interest because the intended semantics are generally relational.)

However, we frequently wish to reason
about programs under certain assumptions about the interaction of atomic
programs and tests.  For example, if
$p$ is the program ``x := 0'' and $b$ is the assertion
``x = 0'', 
then we want to be able to make use of the facts
$pb = p$ (``after running $p$, test $b$ always succeeds'')
and $bp = b$ (``after test $b$ succeeds, $p$ is redundant'')
when reasoning about programs in which $p$ and $b$
appear; 
for instance, the equation $p^2 = p$ is not valid in \KAT,
but the formula
$(pb = p~\wedge~bp = b) \rightarrow p^2 = p$ is.  Thus, 
the \emph{universal Horn theory} is of interest.
A \emph{universal Horn formula} is an implication $E \rightarrow s=t$, where
$E$ is a finite set of equations.  The word ``universal'' refers to the fact
that the atomic symbols of $E$, $s$, and $t$ are implicitly universally
quantified.
The \emph{universal Horn theory} of a class of structures $\C$,
denoted $\Horn{\C}$, is the set
of universal Horn formulas valid under all interpretations over structures
in $\C$.

The increased generality of the universal
Horn theory is accompanied by greater complexity, and the theory does not
remain the same when we restrict to important classes of Kleene algebras
such as $*$-continuous Kleene algebras with tests ($\KAT^*$)
and relational Kleene algebras with tests (\RKAT).  $\Horn{\KAT}$ is
$\Sigma^0_1$-complete (undecidable),
$\Horn{\KAT^*}$ and $\Horn{\RKAT}$ are $\Pi^1_1$-complete (highly
undecidable),
and we have proper inclusions
$\Horn{\KAT} \subsetneq \Horn{\KAT^*}\subsetneq \Horn{\RKAT}$
\cite{Koz02,HK03}. 

Although these Horn theories are very complex in general, there are
fragments of them that are both practical and of lower complexity.
Consider the following theorem, fundamentally due to Cohen \cite{Coh94}
and extended to the form below by Kozen and Smith \cite{KS96, Koz00}.
(The statement uses some notions that will not be defined until later,
but we only need a vague understanding of it here.)
\begin{thm}
\label{thm:subsumehoare}
Let $r,s,t\in\RexPB$, and let $u\in\RexPB$ be the universal regular
expression.  Then the following are equivalent.
\begin{align}
\KAT &\models r = 0 \imp s= t\\
\KAT^* &\models r = 0 \imp s= t\\
\RKAT &\models r = 0 \imp s= t\\
\KAT &\models s + uru = t +uru
\end{align}
\end{thm}
The primary consequence of this theorem is that the Horn theory of Kleene
algebra, restricted to formulas with hypotheses of the form $r=0$, is
decidable, and remains unchanged if we restrict to $*$-continuous or
relational algebras: to decide if $r=0 \imp s=t$ is valid, we simply decide
if $s+uru = t+uru$ is valid.
In this way, we say that we have \emph{eliminated} the hypothesis $r=0$.
It is also possible to eliminate other forms of hypotheses
\cite{Coh94, HK02}.  

The case $r=0$ has particular significance, because partial correctness
assertions can be expressed in $\KAT$ with equations of the form $r=0$
(and multiple equations $r_1=0 \wedge \cdots \wedge r_k = 0$ can be
combined into $r_1+\cdots+r_k = 0$).
So Theorem~\ref{thm:subsumehoare}
shows that the Horn theory of $\KAT$, restricted to hypotheses of the form
$r=0$, subsumes propositional Hoare logic, is decidable, and is furthermore
complete for relational interpretations \cite{Koz00}.

Our main result,
Theorem~\ref{thm:genzeroelim} (p.~\pageref{thm:genzeroelim}),
improves
Theorem~\ref{thm:subsumehoare}
so that $r=0$ can be eliminated in the presence of other hypotheses.
This allows any other technique for eliminating hypotheses to
be extended to include $r=0$.  For example, if we have a technique for
eliminating $f=g$ alone, we can eliminate $f=g \wedge r=0$ by first eliminating
$r=0$ using Theorem~\ref{thm:genzeroelim}, leaving hypothesis $f=g$, which
can then be eliminated.  In this way, Theorem~\ref{thm:genzeroelim} is like
a module for eliminating $r=0$ that can be added on to any other technique
for eliminating hypotheses.

A related result, Corollary~\ref{cor:syntactichomo}, shows that
hypotheses of the form $cp=c$ (where
$c$ is Boolean and $p$ is atomic) can
be eliminated in the presence of other hypotheses, although the
remaining hypotheses are modified.
Hypotheses of the form $cp=c$ are useful for eliminating
redundant code (consider our example $bp = b$ above; it
expresses the fact that $p$ is redundant when $b$ already holds).
(The procedure for eliminating $cp=c$ was introduced in \cite{HK02},
where it was shown how to eliminate $cp=c$ and $r=0$ at the same time.
Without the benefit of Theorem~\ref{thm:genzeroelim}, this required
a construction that simultaneously dealt with both $cp=c$ and $r=0$.)

\section{Preliminaries}

For a more complete introduction to Kleene algebra and Kleene algebra
with tests, see \cite{Koz97}.

\subsection{Kleene Algebra}

\begin{defi}
An \emph{idempotent semiring} is a structure $(S,+,\cdot,0,1)$
satisfying
\begin{align*}
x+x&= x~~~\mbox{(idempotence)}\\
x+0&= x\\
x+y&= y+x\\
x+(y+z)&= (x+y)+z\\
0\cdot x &= x\cdot 0=0\\
1\cdot x &= x\cdot 1=x\\
x\cdot(y\cdot z)&= (x\cdot y)\cdot z\\
x\cdot(y+z)&= x\cdot y + x\cdot z\\
(y+z)\cdot x&= y\cdot x + z\cdot x \dsp.
\end{align*}
(In other words, $(S,+,0)$ is an upper semilattice with bottom element $0$,
$(S,\cdot,1)$ is a monoid, $0$ is an annihilator for $\cdot$,
and $\cdot$ distributes over $+$ on the right
and left.)
\end{defi}

We often write $xy$ for $x\cdot y$.
The upper
semilattice structure induces a natural
partial order on any idempotent semiring:
$x \leq y \Leftrightarrow x +y = y$. 

\begin{defi}
A \emph{Kleene algebra} is a structure
$(K,+,\cdot,^*,0,1)$ such that
$(K,+,\cdot,0,1)$ forms an idempotent semiring, and which satisfies
\begin{align}
1+xx^*&\leq x^*\label{eq:closeleft}\\
1+x^*x&\leq x^*\label{eq:closeright}\\
p+qx \leq x &\imp  q^*p\leq x\label{eq:left}\\
p+xq \leq x &\imp pq^*\leq x\label{eq:right}\dsp.
\end{align}
(The order of precedence among the operators is $^* > \cdot > +$, so that
$p+qr^* = p+(q\cdot(r^*))$.)  We let $\KA$ denote the category of all
Kleene algebras and their homomorphisms.
Equations (\ref{eq:closeleft})--(\ref{eq:right}) are called the
\emph{Kleene algebra $*$-axioms}.

Given a set $\Sigma$ of constant symbols, let
$\Rex_\Sigma$ be the set of Kleene algebra terms over $\Sigma$.
We call the elements of $\Rex_\Sigma$ \emph{regular expressions}, and
the elements of $\Sigma$ \emph{atomic program symbols}.  
An \emph{interpretation} is a homomorphism $I:\Rex_\Sigma \to K$,
where $K$ is a Kleene algebra.  $I$ is determined uniquely by its values
on $\Sigma$.  
\end{defi}

Equations
(\ref{eq:closeleft})  and (\ref{eq:left}) say that
$q^*p$ is the least solution of $p+qx \leq x$,
while
(\ref{eq:closeright}) and (\ref{eq:right}) say that $pq^*$ is the least
solution to $p + xq \leq x$.

A straightforward and vital consequence of the \KA\ axioms%
\footnote{
The names of the categories we consider
serve as convenient abbreviations for the type
of algebra they contain.  So, for example, ``the \KA\ axioms'' means
``the axioms of Kleene algebra''.}
is that
the operations $+$, $\cdot$, and $^*$ are monotone:
if $x_0\leq x_1$ and $y_0\leq y_1$, then $x_0+y_0\leq x_1+y_1$,
$x_0y_0 \leq x_1y_1$, and $x_0^* \leq x_1^*$.


We use $\models$ to denote ordinary Tarskian satisfaction.
However, since we have constant symbols from $\Sigma$ not
in the signatures of the underlying algebras, we will pair each algebra
with an interpretation when speaking about satisfaction.  For example, given
a Kleene algebra $K$, interpretation $I:\Rex_\Sigma \to K$, and
formula $\varphi$ whose atomic program symbols are among
$\Sigma$,
we will write $K,I \models \varphi$ to indicate that $K$ satisfies $\varphi$
when the symbols in $\Sigma$ are evaluated according to $I$.
$K \models \varphi$ means that $K, I \models \varphi$ for every
interpretation $I:\Rex_\Sigma\to K$.  We also use $\models$ in two
other standard ways: for a class $\C$ of algebras,
$\C \models \varphi$ means that $K \models \varphi$ for each
$K \in \C$; for a set $\Phi$ of formulas, $\Phi \models \varphi$
means that $K \models \varphi$ for each algebra $K$ satisfying every formula
in $\Phi$.

We now introduce two particularly important types of Kleene algebras:
\emph{language algebras} and \emph{relational algebras}.

\begin{defi}
\label{defi:candef}
For an arbitrary monoid $M$, its powerset $2^M$ forms a Kleene
algebra as follows.
\begin{eqnarray*}
0 & = &\emptyset\\
1&=&\{1^M\}~~~\mbox{(where $1^M$ is the identity of $M$)}\\
A+B&=&A\cup B\\
A\cdot B &=&\{xy~~|~~x\in A,~y\in B\}\\
A^*&=&\bigcup_{k\in\nats} A^k
\end{eqnarray*}
We let $\REG~M$ denote the smallest subalgebra of $2^M$ containing the
singletons $\{x\}$, $x\in M$.  (The elements of $\REG~M$ are the
\emph{regular subsets} of $M$.)  $2^M$ and its subalgebras are known
as \emph{language algebras}.

Of particular interest is the case $M = \Sigma^*$, 
the monoid of all strings over alphabet $\Sigma$ under
concatenation.  The empty string $\varepsilon$ is the identity of this
monoid.
We define the canonical
interpretation $R:\Rex_\Sigma\to \REG~\Sigma^*$ by
letting $R(p) = \{p\}$ (and extending $R$ homomorphically to the rest 
of $\Rex_\Sigma$).  Note that we can interpret elements of $\Sigma^*$
as elements of $\RexS$ in the obvious fashion.
\end{defi}

\begin{defi}
\label{defi:reldef}
For an arbitrary set
$X$, the set $2^{X \times X}$ of all binary relations on $X$ forms a
Kleene algebra 
as follows.
\begin{eqnarray*}
0 &=& \emptyset\\
1 &=& \id_X = \{(x,x)~~|~~x\in X\}\\
S + T &=& S \cup T\\
S \cdot T &=& S \circ T~~~\mbox{(the composition of $S$ with $T$)}\\
S^* &=& \bigcup_{k\in\nats} S^k~~~\mbox{(the reflexive transitive closure of
$S$)}
\end{eqnarray*}
A Kleene algebra $K$ is \emph{relational} if it is a subalgebra
of $\rel{X}$ for some $X$; $X$ is called the \emph{base}
of $K$.
We let $\RKA$ denote the category of all
relational Kleene algebras and their homomorphisms.
\end{defi}

The definitions of $^*$ in $2^M$ and $\rel{X}$ exemplify the most common
intuition about the meaning of $^*$, which is that
$y^* = \sup_{n \in \nats} y^n$,
or informally, $y^* = 1+y+y^2+\cdots$.  (More generally, if we require
that multiplication distributes over this supremum, we have
$xy^*z = x1z + xyz +xy^2 z + \cdots = \sup_{n\in\nats}xy^n z$.)
However, this property of $^*$
does not follow from the $\KA$ $*$-axioms,
and must be postulated separately.

\begin{defi}
A Kleene algebra $K$ is \emph{$*$-continuous} if it satisfies
$$xy^* z = \sup_{k\in\nats} xy^k z$$
for all $x,y,z\in K$.  We let $\KA^*$ denote the category of all
$*$-continuous Kleene algebras and their homomorphisms.
\end{defi}

Since relational composition distributes over arbitrary union,
it is immediate from the definition of $^*$ in $\rel{X}$ that relational
Kleene algebras are $*$-continuous, so $\RKA \subseteq \KA^*$.

The following ubiquitous lemma is a useful generalization of $*$-continuity.

\begin{lem}
\label{lem:starlem}
Suppose $K\in\KA^*$, $I: \Rex_\Sigma \to K$ is an interpretation, 
and $t\in\Rex_\Sigma$.  Then
$$I(t) = \sup_{\sigma \in R(t)} I(\sigma)\dsp .$$
\end{lem}

\proof
By induction on structure of $t$.  For details,
see \cite[Lemma 7.1, pp. 246--248]{Koz91}.~%
\qed




%


\subsection{Kleene Algebra with Tests}
\label{subsec:kat}

We can combine Kleene algebra with Boolean algebra to get
\emph{Kleene algebra with tests}. 
The Boolean aspect is
useful for capturing Boolean aspects of
programming semantics, particularly control flow and assertions.


\begin{defi}
\label{defi:rkat}
A \emph{Kleene algebra with tests} is a two-sorted structure
$(K,B,+,\cdot,^*,\complementop,0,1)$, where
$(K,+,\cdot,^*,0,1)$ is a Kleene algebra, and
$(B,+,\cdot,\complementop,0,1)$ is a Boolean subalgebra.
The elements of $B$ are called \emph{tests}.
We let $\KAT$ denote the category of all Kleene algebras with tests
and their homomorphisms; we let $\KAT^*$ denote the subcategory of all
$*$-continuous Kleene algebras with tests.

We now have two types of atomic symbols: programs and tests.
For a finite set $\P$ of atomic program symbols
and a finite set $\B$ of atomic test symbols,
$\RexPB$ is the set of $\KAT$ terms over $\P$ and $\B$; negation can only
be applied to Boolean terms, which are terms built from
$0$,$1$,$+$,$\cdot$,$\complementop$, and atomic test symbols.
An interpretation $I:\RexPB \to K$
must map each atomic test to a test in $K$ (and it follows by induction
that it will map all Boolean terms to tests).

$\rel{X}$ forms a Kleene algebra with tests by keeping the previously 
defined Kleene algebra structure, and letting
$B = \{ r \in \rel{X}~~|~~r \leq 1\}$, $\overline{b} = \id_X - b$.
A Kleene algebra with tests $K$ is relational if it is a
subalgebra
of $\rel{X}$ for some $X$.
We let $\RKAT$ denote the category of all relational Kleene algebras with
tests and their homomorphisms.
\end{defi}

Every Kleene algebra induces a Kleene algebra with tests by letting
$B = \{0,1\}$, the two-element Boolean algebra;
conversely, every Kleene algebra with tests induces a Kleene algebra by
taking its reduct to the signature of Kleene algebra
(\emph{i.e.}, taking its image under the map
$(K,B,+,\cdot,^*,\complementop,0,1) \mapsto
(K,+,\cdot,^*,0,1)$).  With this in mind, it is easy to see that for any
formula $\varphi$ in the language of Kleene algebra,
$\KAT \models \varphi \Leftrightarrow \KA \models \varphi$,
$\KAT^* \models \varphi \Leftrightarrow \KA^* \models \varphi$, and
$\RKAT \models \varphi \Leftrightarrow \RKA \models \varphi$.

There is an analog of $\REG~\sigs$ for \KAT\ called the 
\emph{guarded-string model}, with its own analog of the canonical
interpretation $R$.
Though the guarded-string model is in general very important
for studying \KAT, we will not need it for our results here, and refer
the reader to \cite{KS96} for further information on guarded strings.

The following elementary lemma about subalgebras will be needed in
Lemma~\ref{lem:mainlemone}.

\begin{lem}
\label{lem:idealalgebras}
Let $K\in\KA$ and let $x\in K$.  Then 
$\{y\in K~~|~~y\leq x\}$
is a subalgebra of $K$ iff
$x = y^*$ for some $y\in K$ (or equivalently, $x=x^*$).  The same also holds
for $\KAT$s.  (Note that this is not claiming that all subalgebras
of $K$ have this form.)
\end{lem}

The proof is straightforward and may safely be skipped.

\proof\note{Omit proof?}
Let $K' = \{y\in K~~|~~y\leq x\}$.

Suppose $K'$ is a subalgebra of $K$.  Then $x^*\in K'$, so $x^*\leq x$,
so $x = x^*$.

Suppose $x = y^*$ for some $y\in K$.  Then $x^* = y^{**} = y^* = x$.
The necessary closure conditions follow from monotonicity and
the fact that $0 + 1 + xx + (x + x) + x^* \leq x^*$.  (For example,
for any $y_1,y_2\in K'$, we have $y_1 y_2 \leq xx \leq x^*$.)
\qed

\subsection{Universal Horn Formulas}

\begin{defi}
A \emph{universal Horn formula} is a formula of the form
$$s_1 = t_1 \wedge \cdots \wedge s_t = t_k
\imp s = t\dsp,$$
where $s_i, t_i, s, t$ are terms.
The set of universal
Horn formulas valid over a class $\C$ of algebras is the
\emph{universal Horn theory}
of $\C$, which we denote by $\Horn{\C}$.
\end{defi}

We will often drop the word ``universal''.
Note that in \KA\ and \KAT, because any inequality $x\leq y$ is actually
an equation $x+y = y$, inequalities are allowed in Horn formulas.
We will allow 
finite sets of equations to appear in the hypotheses of a Horn
formula, by taking their conjunction; \eg, if $E = \{pq = qp,~p \leq 1\}$,
then $E\imp s=t$ means $(pq =qp \wedge p\leq 1)\imp s=t$.

\begin{lem}
\label{lem:littlestarlem}
Let $\Gamma$ be any class of $*$-continuous Kleene algebras with
interpretations.  (That is, $\Gamma$ consists of pairs
$(K,I)$ where $K\in\KA^*$ and $I:\RexS\to K$ is an interpretation.)
Then for any Horn formula of the form $E\imp s\leq t$,
$$\Gamma\models E\imp s\leq t \Longleftrightarrow
(\forall \sigma\in R(s))~~\Gamma\models E\imp \sigma\leq t\dsp.$$
\end{lem}

\proof
For any $K\in\KA^*$ with interpretation $I:\RexS\to K$, the
equivalence
$$K,I\models E\imp s\leq t
\Longleftrightarrow
(\forall \sigma\in R(s))~~K,I\models E\imp \sigma\leq t$$
is a straightforward consequence of Lemma~\ref{lem:starlem}.
The lemma then
follows by exchanging the universal quantifiers
$(\forall \sigma\in R(s))$ and $(\forall (K,I)\in\Gamma)$.  (This latter
quantifier comes from $\Gamma\models E\imp s\leq t
\Leftrightarrow (\forall (K,I)\in\Gamma)~K,I\models E\imp s\leq t$.)
\qed

%
%
%
%

\subsection{A Proof System for \Horn{\RKA}}
\label{sec:proofsystem}

Later, in the proof of Lemma~\ref{lem:mainlemthree}, we will use a
proof-theoretic argument based on the infinitary proof system
for $\Horn{\RKA}$ introduced in \cite{Hard05proof}.  We will only present
the material that we will need in Section~\ref{sec:genzeroelim}
for the proof of Lemma~\ref{lem:mainlemthree};
for a more thorough treatment, please see \cite{Hard05proof}.

\longnote{
The definition of $R(s)$ when $s$ has tests is handled
as in the LICS abstract: we assume WLOG that only atomic tests are
negated, and for any atomic test $b$, we treat $b$ and
$\overline{b}$ as atomic programs, and pretend that any Horn formulas
have the additional hypotheses $b\overline{b} = 0$ and $b+\overline{b}=0$.
(Admittedly, it would be nice to refine the proof system so that tests are
more naturally incorporated.)
If you are uncomfortable with the use of tests in the proof
of (\ref{eq:relimthree}) above,
we can just use the above proof with $\RKA$ in
place of $\RKAT$, and use the following lemma to extend the result to
$\RKAT$.
(The translation defined in the following lemma is
essentially the above process: replace Boolean terms with
terms in which only atomic tests are negated, treat $b$ and
$\overline{b}$ as atomic programs, and add hypotheses $b+\overline{b} = 1$
and $b\overline{b} = 0$.)
}

\subsubsection{Finite Automata and Trees}

Our proof system for $\Horn{\RKA}$ is based
on trees of finite automata, and we must define a number of notions related
to trees and automata before continuing.

Assume we have a fixed finite alphabet $\Sigma$.
We let \nfa\ denote 
the set of all nondeterministic finite automata over $\Sigma$,
allowing $\varepsilon$-moves (also called $\varepsilon$-edges).

We will also use NFA as shorthand
for \emph{nondeterministic finite automaton}.  For any NFA $A$, $L(A)$ denotes
the language of $A$, and $|A|$ denotes the states of $A$.
For states $v,w\in |A|$,
let $A^{v,w}$ denote the NFA which is
identical to $A$ except that it has $v$ and $w$ as its unique start and accept
states, respectively.
We fix distinct states $a$ and $b$,
and let $\nfa^{a,b}$ be the set of all $A\in\nfa$
which have unique start state $a$ and unique accept state $b$.




We define $F_0\in \nfa^{a,b}$ to have states
$\{a,b\}$ and no edges.

Given an NFA $A$ and states $v,w\in |A|$, we will sometimes want to
``insert'' a string $\tau\in \Sigma^*$ into $L(A^{v,w})$.  For this
purpose, we define $A' = \ins_2(A,v,w,\tau)$ as follows.
\begin{enumerate}
\item If $\tau=p_1\cdots p_k$, with $p_i\in \Sigma$ and $k > 0$, we obtain $A'$
from $A$ by adding $k-1$
new states $x_1,\ldots,x_{k-1}$ and adding edges
$$v
\stackrel{p_1}{\rightarrow} x_1
\stackrel{p_2}{\rightarrow} \cdots
\stackrel{p_{k-1}}{\rightarrow} x_{k-1}
\stackrel{p_{k}}{\rightarrow} w\dsp.$$

\item If $\tau = \varepsilon$, then we add an $\varepsilon$-edge from
$v$ to $w$
and also from $w$ to $v$.
(Where it is used,
$\ins_2(A, v,w,\varepsilon)$
corresponds to identifying $v$ and $w$ with each other.
The edge from $w$ to $v$,
called a \emph{reverse $\varepsilon$-edge},
is needed to capture the
symmetry of the identity relation.)
\end{enumerate}


We now move on to trees.
$\nats^*$ is the set of all finite strings of naturals (including
the empty string).
A set $T \subseteq \nats^*$ is a \emph{tree} if it is closed under
taking initial segments.
A function $f:\nats\to \nats$ can be treated as an infinite
sequence of naturals, and for $n\in\nats$, we let $f\restriction n$ denote
the initial segment of $f$ of length $n$.
Such an $f$ is a \emph{path} through a tree $T$ if
$(f\restriction n )\in T$ for all $n \in\nats$.
(We find this a concise framework for countably-branching trees, but it is
not strictly necessary to define trees in this manner.)

\subsubsection{Relational Proofs}

The following definition of \emph{relational proof} captures, with
trees of finite automata, the combinatorics of attempting to construct a
relational counterexample to a Horn formula.
A path through such a tree yields a relational model in which
the formula fails, while
well-foundedness establishes the impossibility of a counterexample
({\it i.e.}, the relational validity of the formula).

\begin{defi}
\label{defi:relproof}
Let $E \imp \sigma\leq t$ be a
Horn formula in the language of $\KA$ with $\sigma \in\Sigma^*$ and
$t\in\RexS$.  We assume that
all hypotheses in $E$ are inequalities $x \leq y$, by breaking any equations
$x = y$ into $x \leq y \wedge y \leq x$ as necessary.  
We fix distinct states $a$ and $b$ as above.
We fix a special symbol \con, which will signify
contradiction.

A \emph{relational tree for $E \imp \sigma \leq t$} is a pair
$(T,A)$ where $T \subseteq \nats^*$ is a tree and
$A:T \to \nfa^{a,b} \cup \{\con\}$ such that the following
conditions hold.  ($A_f$ will denote $A(f)$.)
\begin{enumerate}
\item At the root, we have 
$A_{\langle\rangle} = \ins_2(F_0,a,b,\sigma)$.


\item $f\in T$ is a leaf node if and only if $A_f = \con$ or 
$R(t)\cap L(A_f)\neq \emptyset$.

\item If $f$ is not a leaf node, then
there exist $v,w\in |A_f|$
(possibly equal),
an inequality $r\leq r'$ in $E$,
and $\rho \in L(A_f^{v,w}) \cap R(r)$ such that
\begin{enumerate}
\item if $R(r') = \emptyset$ (typically because $r' = 0$), then $f$
has one child $g$, with $A_g = \con$;
\item if $R(r') \neq \emptyset$, then $f$ has one child $g_\tau$ for each
$\tau \in R(r')$, with
$A_{g_\tau} = \linebreak \ins_2 (A_f, v,w, \tau)$.
\end{enumerate}
(We say that the hypothesis $r\leq r'$ is \emph{applied at $f$}.)
\end{enumerate}

A \emph{relational proof of $E\imp \sigma\leq t$} is a well-founded
relational tree for $E\imp \sigma \leq t$.
We say $E\imp \sigma\leq t$
is \emph{relationally provable} if such a proof exists.
\end{defi}

\begin{lem}
\label{lem:mainrkalem}
For any Horn formula of the form $E \imp \sigma \leq t$,
the following are equivalent.
\begin{enumerate}
\romanize
\item $\RKA \models E \imp \sigma \leq t$
\item $E \imp \sigma \leq t$ is relationally provable.
\end{enumerate}
\end{lem}

\proof
See \cite{Hard05thesis} or \cite{Hard05proof}.
\mynote{Careful here: the proof only appears in the appendix of
\cite{Hard05proof}, which does not appear in the proceedings version.}
\qed

The notion of relational provability can be extended
to arbitrary Horn formulas, but we will not need it for the proof
of Lemma~\ref{lem:mainlemthree}.




\subsection{The Relationship Between \Horn{\RKA} and \Horn{\RKAT}}
\label{sec:relationship}

The system presented in Section~\ref{sec:proofsystem} is a tool
for studying \Horn{\RKA}, while in Lemma~\ref{lem:mainlemthree}, we
will wish to use it to draw conclusions about \Horn{\RKAT}.
This must be rectified,
and there are multiple ways to proceed.  One would be to modify the
notion of relational proof so that it applies to \Horn{\RKAT};
this would present no
particular difficulty, but would require a closer look at relational proofs
than we would like to get into here.  Instead, we will show how to
reduce questions about \Horn{\RKAT}\ to \Horn{\RKA}, in a way that will
allow us to use the existing definition of relational proof
when proving Lemma~\ref{lem:mainlemthree}.

\begin{lem}
\label{lem:trlem}
For any Horn formula $\varphi$ of \KAT, there is a Horn
formula $\Tr(\varphi)$ of \KA\ such that
$\RKAT \models \varphi$ iff $\RKA \models \Tr(\varphi)$. 
\end{lem}

The lemma is uninteresting without putting restrictions on
the translation $\Tr$.  However, instead of trying to
capture the desired properties of $\Tr$ for inclusion in the lemma,
we just give the proof, and observe
later that the translation works for a particular purpose when the need
arises.  \note{(Informally, the property of $\Tr$ that we will need is that it
commutes with certain other syntactic operations.)}

\proof
(Outline: we first assume that negation is only applied to atomic
tests, then replace the negations of atomic tests with fresh program symbols,
and finally
add new hypotheses to ensure that the new program symbols behave like
the negated tests they replace.)

Fix a set $\P$ of atomic program symbols, and a set $\B$ of atomic tests.
Given any $s\in\RexPB$, we can assume without loss of generality that
negation is only applied to atomic tests, in light of DeMorgan's 
Laws\note{Law?}.

For each $b\in\B$, we introduce two new atomic program symbols
$\tilde{b}$
and
$\tilde{\overline{b}}$, and we let
$\Sigma = \P \bigcup \{\tilde{b},\tilde{\overline{b}}~~|~~b\in\B\}$.  For
any $t\in\RexPB$, we let $\tilde{t}$ be the result of taking $t$, 
and replacing all occurrences of $\overline{b}$ with $\tilde{\overline{b}}$,
and all positive occurrences of $b$ with $\tilde{b}$ (for each $b\in\B$).
Note that $\tilde{t}\in\Rex_{\Sigma}$. For any formula $\varphi$, we let
$\tilde{\varphi}$ be the result of replacing each term $t$ in $\varphi$
with $\tilde{t}$.

Now take any Horn formula $\varphi$ of the form
$\theta \imp \psi$ (with
all terms in $\RexPB$).
Let $\Tr(\varphi)$ be the formula
$$\left(\tilde{\theta} \wedge
\bigwedge_{b\in\B}
(\tilde{b}+\tilde{\overline{b}} = 1 \wedge
\tilde{b}\cdot\tilde{\overline{b}} = 0) \right)
\imp \tilde{\psi}\dsp.$$
(The extra hypotheses make $\tilde{b}$ and $\tilde{\overline{b}}$
behave like Boolean complements of each other.)

We now show 
$\RKAT \models \varphi$ iff $\RKA \models \Tr(\varphi)$.

For the right-to-left implication,
suppose $\RKAT \not \models \varphi$.  Let $K\in\RKAT$ with
interpretation $I:\RexPB\imp K$
such that $K,I\not\models \varphi$.  Then
$K,I\models\theta \wedge \neg\psi$.
Define the interpretation $\tilde{I}: \Rex_\Sigma \to K$ by
$$\tilde{I}(p) = \left\{
\begin{array}{ll}
I(p), & \mbox{if $p \in \P$,}\\
I(b), &\mbox{if $p=\tilde{b}$,}\\
I(\overline{b}), &\mbox{if $p=\tilde{\overline{b}}$.}
\end{array}\right.$$

A simple induction shows that for any $t\in\RexPB$,
$\tilde{I}(\tilde{t}\,) =
I(t)$.  It follows that
$K,\tilde{I}\models \tilde{\theta}\wedge\neg\tilde{\psi}$, since
$K,I\models \theta\wedge\neg\psi$.  Also,
$$K,\tilde{I}\models\bigwedge_{b\in\B}
(\tilde{b}+\tilde{\overline{b}} = 1 \wedge
\tilde{b}\cdot\tilde{\overline{b}} = 0)\dsp.$$
Thus $K,\tilde{I}\not\models \Tr(\varphi)$, so
$\RKA \not\models\Tr(\varphi)$ (recall that we can treat $K$ as a member
of \RKA\ by passing it through the forgetful functor which drops negation).
Therefore, $\RKA \models \Tr(\varphi) \imp \RKAT \models \varphi$.

For the left-to-right implication,
suppose that $\RKA \not\models\Tr(\varphi)$.  Let $K\in\RKA$ with
interpretation $I:\Rex_{\Sigma}\to K$
such that $K,I\not\models\Tr(\varphi)$.  Let $X$ be the base of $K$.
Then $K \subseteq \rel{X}$, so 
$\rel{X}, I \not\models\Tr(\varphi)$; that is,
$$\rel{X},I\models
\tilde{\theta} \wedge
\bigwedge_{b\in\B}
(\tilde{b}+\tilde{\overline{b}} = 1 \wedge
\tilde{b}\cdot\tilde{\overline{b}} = 0)
\wedge\neg\tilde{\psi}\dsp.$$
In particular, for any $b\in\B$,
$I(\tilde{b}) \bigcup I(\tilde{\overline{b}}) = I(1)$, and
$I(\tilde{b}) \circ I(\tilde{\overline{b}}) = \emptyset$; it follows that
$I(\tilde{b}) \bigcap I(\tilde{\overline{b}}) = \emptyset$ (since
$R \bigcap S = R \circ S$ whenever $R,S \subseteq I(1)$), so
$I(\tilde{\overline{b}}) = I(1) - I(\tilde{b})$.

Define the interpretation $I':\RexPB\to \rel{X}$ by
\begin{eqnarray*}
I'(p)&=&I(p)\dsp,\\
I'(b)&=&I(\tilde{b})\dsp.
\end{eqnarray*}
We have
\begin{eqnarray*}
I'(\overline{b}) &= &I'(1) - I'(b)\\
&=& I(1) - I(\tilde{b})\\
&=& I(\tilde{\overline{b}})\dsp.
\end{eqnarray*}
It follows that, for any $t\in\RexPB$, $I'(t) = I(\tilde{t}\,)$.
So, $\rel{X}, I' \models \theta \wedge\neg\psi$, since
$\rel{X}, I \models \tilde{\theta} \wedge\neg\tilde{\psi}$, giving us
$\RKAT \not\models \varphi$.
Therefore,
$\RKAT \models \varphi
\imp
\RKA \models \Tr(\varphi)$, completing the proof.
\qed


\section{Main Results}
\label{sec:mainresults}

\subsection{Eliminating $r=0$}
\label{sec:genzeroelim}

\begin{defi}
For a fixed set $\P = \{p_1,\ldots,p_n\}$ of atomic program symbols,
the \emph{universal regular expression $u$} is defined by
$$u = (p_1+\cdots+p_n)^*\dsp.$$
\end{defi}

We trivially have $\KAT \models u = uu = u^*$, and a straightforward
induction shows that, for any $s\in\RexPB$, $\KAT \models s \leq u$.

Our goal is the following theorem.

\begin{thm}
\label{thm:genzeroelim}
Let $u$ be the universal regular expression,
let $E$ be any finite set of hypotheses, and let $r,s,t\in\RexPB$.
Then the following equivalences hold.
\begin{alignat}{2}
\KAT \models E\wedge r=0 \imp s=t
&\iff &
\KAT \models E\imp s+uru=t+uru\label{eq:relimone}\\
\KAT^* \models E\wedge r=0 \imp s=t
&\iff &
\KAT^* \models E\imp s+uru=t+uru\label{eq:relimtwo}\\
\RKAT \models E\wedge r=0 \imp s=t
&\iff&
\RKAT \models E\imp s+uru=t+uru\label{eq:relimthree}
\end{alignat}
\end{thm}

Note that the special case $E = \emptyset$ is essentially
Theorem~\ref{thm:subsumehoare} (when $E = \emptyset$, the right
hand sides of \eqref{eq:relimone}--\eqref{eq:relimthree}
are equivalent, since the equational theories of $\KAT$,
$\KAT^*$, and $\RKAT$ coincide; when $E\neq \emptyset$, the right hand sides
of \eqref{eq:relimone}--\eqref{eq:relimthree} are no longer necessarily
equivalent, which prevents Theorem~\ref{thm:genzeroelim} from having the
same form as Theorem~\ref{thm:subsumehoare}).
Note also that for any formula $\varphi$ in the language of $\KA$,
we have $\KA\models \varphi$ iff $\KAT\models\varphi$,
$\KA^*\models \varphi$ iff $\KAT^* \models\varphi$, {\it etc.}, so
Theorem~\ref{thm:genzeroelim} also applies to $\KA$, $\KA^*$, and $\RKA$.
(Alternatively, omitting the Boolean aspects of the proof that
follows would yield
a proof of the analogous theorem for $\KA$, $\KA^*$, and $\RKA$.)

We prove each equivalence as a separate lemma.  Fix $u$, $E$,
$r$, $s$, $t$,
as above.

\begin{lem}
\label{lem:mainlemone}
$$\KAT \models E\wedge r=0 \imp s=t
\iff
\KAT \models E\imp s+uru=t+uru$$
\end{lem}

\proof
The right-to-left implication is trivial:
reasoning under $E\wedge r=0$,
we have $s = s+0 = s+uru= t+uru= t+0=t$.  (Note that this
argument also applies to $\KAT^*$ and $\RKAT$.)

For the left-to-right implication,
suppose
$\KAT\models E\wedge r=0\imp s=t$.
Take any $K\in\KAT$ with interpretation $I$ such that
$K,I\models E$.  
Let $\bot = I(uru)$, $\top = I(u)$, noting that
$\top^* = \top$, $\bot= \top\bot = \bot\top$, and
$\bot\bot\leq\bot$.
Let $K' = \{x\in K~~|~~x\leq \top\}$.
This is a subalgebra of $K$ by
Lemma~\ref{lem:idealalgebras},
since $\top = \top^*$.
$I$ is an interpretation into $K'$.  

Define the map $f:K'\rightarrow K'$ by $f(x) = x+\bot$.
Let $L = f[K']$, the image of $K'$ under $f$.
$\top$ and $\bot$ are respectively the
greatest and least elements of $L$.
Note that for any $x\in K'$, $x\top\leq \top$, so
$x\bot = x\top\bot \leq \top\bot = \bot$.  We similarly have
$\bot x \leq \bot$.

Define 
\begin{eqnarray*}
0^L &=& \bot ~~=~~ f(0)\\
1^L &=& 1 + \bot ~~=~~ f(1)\\
v\cdot^L w &=& v\cdot w + \bot = f(vw)\dsp.
\end{eqnarray*}

Let $L$ be the structure
$(L,f[B],+,\cdot^L,{}^*,\tilde{\enspace},0^L,1^L)$, 
in the signature of \KAT,
where
$B$ is the set of tests of $K'$, and the 
Boolean complement $\tilde{\enspace}$ is defined by
$\widetilde{f(c)} = f(\overline{c})$.  We must show that
$\tilde{\enspace}$ is well-defined.
Suppose $f(c) = f(d)$.
Then
\begin{eqnarray*}
f(\overline{c}) &=& \overline{c}+\bot\\
&\leq& (\overline{c}+\bot)(1+\bot)\\
&=& (\overline{c}+\bot)(d +\overline{d} + \bot)\\
&=& (\overline{c}+\bot)(c +\overline{d} + \bot)\quad\mbox{(since
$c+\bot =f(c)=f(d)= d+\bot$)}\\
&=& \overline{c}c + \overline{c}\overline{d} + \overline{c}\bot
+\bot c + \bot \overline{d} +\bot\bot\\
&\leq& 0+\overline{d}+\bot\\
&=&f(\overline{d})\dsp.
\end{eqnarray*}
Similarly, $f(\overline{d})\leq f(\overline{c})$,
so $f(\overline{c})=f(\overline{d})$.  Therefore, 
$\tilde{\enspace}$ is well-defined.

We claim that $f:K'\rightarrow L$ is a homomorpishm.  (Note that
this is different from claiming that $f:K'\rightarrow K'$ is a homomorphism,
which is not true unless $\bot = 0$.)
For any $x,y\in K$, and $c$ a test in $K$,
\begin{eqnarray*}
f(0) &=& 0^L\\
f(1) &=& 1^L\\
f(x+y) &=& x+y+\bot = x+\bot+y+\bot = f(x) + f(y)\\
f(xy) &=&xy+\bot\\
     & =& xy + \bot y + x \bot +\bot\bot+ \bot
\quad\mbox{(since $\bot y + x\bot+\bot\bot \leq \bot$)}\\
&=& (x+\bot)(y+\bot)+\bot\\
&=& f(x)\cdot^L f(y)\\
f(\overline{c}) &=&\widetilde{f(c)}\dsp.
\end{eqnarray*}
It remains to verify $f(x^*) = (f(x))^*$.
We have
$$1 + (x+\bot)(x^*+\bot) = 1 + xx^* + x\bot + \bot x^* +\bot\bot
\leq  x^* + \bot\dsp,$$
so the $*$-axioms give us $(x+\bot)^* \leq x^* + \bot$.
We have $x^* \leq (x+\bot)^*$ and $\bot \leq (x+\bot)^*$ trivially,
so
$x^* + \bot \leq (x+\bot)^*$.
Therefore, $f(x^*) = x^*+\bot = (x+\bot)^* = (f(x))^*$.
So $f:K'\rightarrow L$ is a homomorphism.

We now claim that $L\in\KAT$.  Since $f:K'\rightarrow L$ is a homomorphism
and $K'\in\KAT$,
$L$ automatically satisfies the
equational $\KAT$ axioms. We must now verify that $L$ satisfies the
two remaining axioms, $p+q\cdot^L x \leq x \imp q^* \cdot^L p \leq x$
and $p+x\cdot^L q \leq x \imp p\cdot^L q^* \leq x$.

Suppose that $p+q\cdot^L x \leq x$.  We must show $q^* \cdot^L p \leq x$.
We have $p+qx+\bot = p+q\cdot^L x \leq x$.  
From $p+qx\leq x$ we conclude
$q^* p \leq x$; combining this with $\bot \leq x$, we have
$q^*\cdot^L p = q^* p + \bot \leq x$, as desired.
Similarly, $p+x\cdot^L q \leq x \imp p\cdot^L q^* \leq x$.
So $L\in\KAT$.

Define the interpretation $J:\RexPB\rightarrow L$
by $J(q) = f(I(q))$.
Since $K',I\models E$, it immediately follows
that $L,J\models E$.  Also,
$J(r) \leq J(uru) = f(I(uru)) = f(\bot) = \bot+\bot = 0^L$, so
$L,J\models r=0$.  Therefore, the assumption
$\KAT\models E\wedge r=0 \imp s=t$ gives us
$L,J\models s=t$.
Therefore,
$$I(s+uru) = I(s)+I(uru) = I(s) +\bot = f(I(s)) = J(s) = J(t) =
I(t+uru)\dsp.$$
Thus, $K,I\models s+uru=t+uru$.
\qed

\begin{lem}
\label{lem:mainlemtwo}
$$\KAT^* \models E\wedge r=0 \imp s=t
\iff
\KAT^* \models E\imp s+uru=t+uru$$
\end{lem}

\proof
The right-to-left implication is as in Lemma~\ref{lem:mainlemone}.

For the left-to-right implication, it suffices to verify that
the construction in the proof of Lemma~\ref{lem:mainlemone}
preserves $*$-continuity.  Letting $q^{(n)}$ denote the $n^\mathrm{th}$
power of $q$ under $\cdot^L$ (with $q^{(0)} = 1^L$), we have
\begin{align*}
\sup_n p\cdot^L q^{(n)} \cdot^L r & = 
\sup_n (pq^n r + \bot)\\
&= pq^* r + \bot\\
&= p\cdot^L q^* \cdot^L r\dsp.
\end{align*}
(For the second equality above, one can observe that
$pq^n r + \bot \leq pq^* r + \bot$ for all $n$, and that
if $x$ is any upper bound for $pq^n r + \bot$, then 
$pq^* r = \sup_n pq^n r \leq x$ and $\bot \leq x$, so
$pq^* r + \bot \leq x$.  So $\sup_n (pq^n r + \bot) = pq^* r + \bot$.)
\qed

\begin{lem}
\label{lem:mainlemthree}
$$\RKAT \models E\wedge r=0 \imp s=t
\iff
\RKAT \models E\imp s+uru=t+uru$$
\end{lem}

\proof
The right-to-left implication is as in
Lemma~\ref{lem:mainlemone}.

For the left-to-right implication,
using the above construction would require
verifying that $L$ has a relational representation, which is not clear.
Instead,
we use a proof-theoretic argument.
Suppose
$\RKAT\models E\wedge r=0\imp \sigma\leq t$, where $\sigma\in R(s)$.
$r=0$ is equivalent to $r\leq 0$, and $\KAT\models t\leq t+uru$, so 
$\RKAT\models E\wedge r\leq 0\imp \sigma\leq t+uru$.

For the moment, suppose that the formulas are in the language of \KA, so
that we can speak about relational proofs without worrying about tests.
Let $(T,A)$ be a relational proof of
$E\wedge r\leq 0 \imp \sigma\leq t+uru$.

We claim that the hypothesis $r\leq 0$ is never even applied in the proof!
Suppose $r\leq 0$ is applied at node $f\in T$ (so 
$f$ has
one child $g$ with $A_g = \con$).
For $r\leq 0$ to be applied at $f$,
there must be states $v,w\in |A_f|$ and $\rho\in R(r)$ with
$\rho\in L(A_f^{v,w})$.
A property that is preserved in the automata of relational
trees is that every state is accessible from the start state $a$,
and the accept state $b$ is accessible from every state.  So
there exist $\pi\in L(A_f^{a,v})$ and $\pi' \in L(A_f^{w,b})$.  Thus,
we have $\pi\rho\pi'\in 
L(A_f)$;
we also have $\pi\rho\pi'\in R(uru) \subseteq R(t+uru)$.
Therefore, $R(t+uru)\cap L(A_f)\neq \emptyset$, so $f$ is in fact a leaf
node, contradicting the assumption that we are applying $r\leq 0$ at $f$.
(In other words, at any point in a relational tree for
$E\wedge r \leq 0 \imp \sigma \leq t+uru$ where we could apply
$r\leq 0$, we would already have to be at a leaf.)

So, because $r\leq 0$ is never applied,
$(T,A)$ is also a relational proof of
$E\imp \sigma \leq t+uru$.
Therefore, $\RKA\models E\imp\sigma \leq t+uru$ for all
$\sigma\in R(s)$.  By Lemma~\ref{lem:littlestarlem},
$\RKA\models E\imp s \leq t+uru$, so
$\RKA\models E\imp s+uru \leq t+uru$.
$\RKA\models E\imp t+uru \leq s+uru$ is similar, and we now have
$\RKA\models E\imp s+uru = t+uru$.

In case the formulas are not in the language of \KA, we can use
the translation from Section~\ref{sec:relationship} as follows.
We use the above argument to get
\[
\RKA\models\Tr(E\wedge r=0 \imp s=t)
\Rightarrow
\RKA\models\Tr(E\imp s+uru=t+uru)\dsp.\]
(The extra hypotheses introduced by the translation may be
treated the same
as the hypotheses in $E$.  
A subtle point here is that the translation introduces new program symbols,
without adding them to the universal regular expression; however,
the hypotheses added by the tranlation force the interpretations of these
extra symbols to be below 1, so they could be added to the universal
regular expression without affecting the validity of any formulas involved.)
We then have
\begin{align*}
\RKAT\models E\wedge r=0 \imp s=t &\Rightarrow
\RKA\models\Tr(E\wedge r=0 \imp s=t)\\
&\Rightarrow \RKA\models\Tr(E\imp s+uru=t+uru)\\
&\Rightarrow \RKAT\models E\imp s+uru=t+uru\dsp.
\end{align*}
\vspace{-0.42in}

\qed

\proof[Proof of Theorem~\ref{thm:genzeroelim}]
Immediate from Lemmas~\ref{lem:mainlemone}--\ref{lem:mainlemthree}.
\qed


\subsection{Idempotent Syntactic Homomorphisms}

We can also eliminate hypotheses of the form $cp = c$ ($c$ Boolean, $p$
atomic) in the presence of other hypotheses, but not as cleanly as
we eliminated $r=0$:
in this case, the remaining hypotheses
will be modified.

The basic idea behind the technique was introduced in \cite{HK02},
which showed how to simultaneously eliminate hypotheses of
the form $cp = c$ and $r=0$.
Ernie Cohen later observed that
the portion of the proof specific to $cp=c$ was unnecessarily complicated
\cite{ErnieCohenEmail}.
What we present here is a simplified argument, that is also
more general because it works in the presence of other hypotheses.
Furthermore, in light of Theorem~\ref{thm:genzeroelim},
we no longer
need to worry about integrating the elimination of $r=0$ into
the argument, since that can be done separately.

\begin{defi}
\label{defi:syntactichomo}
$H : \RexPB \to \RexPB$ is a \emph{syntactic homomorphism}
if for any 
interpretation $I:\RexPB\to K$ (where $K\in\KAT$),
$I\circ H :\RexPB \to K$ is also an interpretation.
\note{In thesis, write this up in terms of guarded-string interpretation.}

For any syntactic homomorphism $H : \RexPB \to \RexPB$,
let $E_H$ be the set of hypotheses
\[\{p = H(p)~~|~~p\in\P\}\union\{b = H(b)~~|~~b\in\B\}\dsp.\]
\end{defi}

Definition~\ref{defi:syntactichomo} is equivalent to saying that
$H$ is a homomorphism up to $\KAT$-provable equality.  A
consequence is that $H$ is uniquely determined (up to $\KAT$-provable
equality) by its action on $\P$ and $\B$; the set of equations $E_H$ then,
in a certain sense, captures the action of $H$.

(For readers familiar with
guarded strings, Definition~\ref{defi:syntactichomo} is equivalent to
saying that $G\circ H$ is an interpretation, where $G$ is the
guarded-string interpretation.  More abstractly, the definition is equivalent
to saying that $H$ is a lift of an endomorphism on the guarded-string
model---that is, there is an endomorphism $h$ on the guarded-string model
such that $G\circ H = h\circ G$.)

\begin{lem}
\label{lem:syntactichomo}
If $H:\RexPB\to\RexPB$ is a syntactic homomorphism, then for any
$r\in\RexPB$,
\[\KAT\models E_H\rightarrow r = H(r)\dsp.\]
\end{lem}

\proof
Straightforward induction on the structure of $r$.
\note{More detail?}
\qed

\begin{defi}
\label{defi:idempotent}
$H :\RexPB \to \RexPB$ is \emph{idempotent} if for all $r\in\RexPB$,
\[\KAT\models H(r) = H(H(r))\dsp.\]
\end{defi}

\begin{thm}
\label{thm:syntactichomo}
Suppose $H:\RexPB\to\RexPB$ is an idempotent syntactic homomorphism,
and that $E$ is a set of hypotheses.  Let $H(E)$ denote
the set of hypotheses
\[\{H(r) = H(r')~~|~~\mbox{$r=r'$ is in $E$}\}\dsp.\]  Then for any
$s,t\in\RexPB$ and $K\in\KAT$,
\[K\models E \wedge E_H \imp s=t
\iff K\models H(E) \imp H(s) = H(t)\dsp.\]
\end{thm}

\proof
For the right-to-left implication, suppose
$K\models H(E) \imp H(s) = H(t)$ and that 
we have an intepretation
$I:\RexPB\to K$ with $K,I\models E\wedge E_H$.
Then by Lemma~\ref{lem:syntactichomo}, $K,I \models H(E) \wedge
s = H(s) \wedge t = H(t)$.  It follows by assumption that
$K,I\models H(s) = H(t)$.  We now have
$K,I\models s = H(s) = H(t) = t$.  Therefore,
$K\models E\wedge E_H \imp s=t$.

For the left-to-right implication,
suppose $K\models E\wedge E_H \imp s=t$,
and that 
we have an intepretation
$I:\RexPB\to K$ with $K,I\models H(E)$.
Define $I':\RexPB\to K$ by $I' = I \circ H$. $I'$ is an interpretation
by Definition~\ref{defi:syntactichomo}.  For any $p\in\P$,
idempotence of $H$ gives us $I'(p) = I(H(p)) = I(H(H(p))) = I'(H(p))$;
similarly, $I'(b) = I'(H(b))$ for $b\in\B$, so $K,I'\models E_H$.
For any equation $r = r'$ in $E$, $K,I\models H(E)$ gives us
$I'(r) = I(H(r)) = I(H(r')) = I'(r')$, so $K,I' \models E$.
Therefore, by the assumption $K\models E\wedge E_H\imp s=t$,
we have $K,I' \models s=t$, and hence $I(H(s)) = I'(s) = I'(t) = I(H(t))$.
Therefore $K,I \models H(s) = H(t)$, as desired.
\qed

\begin{cor}
\label{cor:syntactichomo}
Suppose $F$ is a set of hypotheses $c_i p_i = c_i$, $1 \leq i \leq k$,
where $p_i \in \P$ are distinct, and each $c_i$ is a Boolean term.
Define $H:\RexPB \to\RexPB$ by $H(r) = r[p_i/\overline{c_i}p_i + c_i]$,
the result of substituting $\overline{c_i}p_i + c_i$ for $p_i$
in $r$ (for each $i$).  Then for any set $E$ of hypotheses,
$s,t\in\RexPB$, and $K\in\KAT$, we have
\[K\models E\wedge F \rightarrow s=t
\iff
K\models H(E) \rightarrow H(s) = H(t)\dsp.\]
\end{cor}

\proof
It is easy to verify that $H$ is an idempotent syntactic homomorphism.

Next, observe that $\KAT \models c_i p_i = c_i \leftrightarrow
p_i = \overline{c_i}p_i + c_i$.
Every equation in $E_H$ is either of the
form $p_i = \overline{c_i}p_i + c_i$, or is a tautology such as $b=b$,
so $F$ is equivalent to $E_H$.  
The corollary now follows immediately from
Theorem~\ref{thm:syntactichomo}.
\qed


The restriction that the $p_i$ be distinct in
Corollary~\ref{cor:syntactichomo} is not a significant imposition,
since we can combine $c_i p_i = c_i$ and $c_j p_j = c_j$, for $p_i = p_j$,
into $(c_i + c_j)p_i = c_i+c_j$.
(Supposing $cp=c$ and $dp = d$, we have $(c+d)p = cp + dp = c+d$.
Supposing $(c+d)p = c + d$, we have $c\leq c+d$, so $c(c+d) = c$,
giving us $cp = c(c+d)p = c(c+d) = c$; $dp = d$ follows similarly.)


\section{Conclusion and Further Questions}

Statements about the semantics of a program can often be expressed as
Horn formulas in Kleene algebra with tests, and that is our primary
motivation for studying the Horn theory of Kleene algebra with tests here.
Hypotheses of the form $r=0$ are of particular interest, because they
can capture partial correctness assertions, which are vital to studying the
semantics of imperative programs.

While the validity of Horn formulas in Kleene algebra is not in
general decidable, the validity of equations is.  We have shown how to
eliminate hypotheses of the form $r=0$, even in the presence of other
hypotheses; this allows us to extend any other technique for
eliminating hypotheses to include hypotheses of the form $r=0$.  We
have also shown how to eliminate hypotheses of the form $cp=c$ in the
presence of other hypotheses (though not as cleanly: the remaining
hypotheses might be modified).  This allows us to decide the validity
of Horn formulas that have hypotheses of these forms.

The following are a few questions for further work.
What other forms of hypotheses can be eliminated?  Can they be
eliminated in the presence of other hypotheses?  Are there useful
decision procedures for the validity of certain classes of Horn
formulas that are not based on eliminating hypotheses?


\section{Acknowledgments}

This work was supported in part by NSF grant CCR-0105586 and by ONR Grant
N00014-01-1-0968.  The views and conclusions contained herein are those of
the author and should not be interpreted as necessarily representing the
official policies or endorsements, either expressed or implied, of these
organizations or the US Government.

\bibliographystyle{plain}

\end{document}